 \documentstyle[twoside,fleqn,espcrc2,psfig,epsfig]{article}

\newcommand{\AmS}{{\protect\the\textfont2
  A\kern-.1667em\lower.5ex\hbox{M}\kern-.125emS}}

\title{LMA parameters and non-zero $U_{e3}$ effects on atmospheric
$\nu$ data? }

\author{O. L. G. Peres\address[MCSD]{Instituto de Fisica Gleb
Wataghin, Universidade Estadual 
de Campinas, UNICAMP 13083-970 Campinas SP, Brazil}%
        \thanks{O.L.G.P. thanks the hospitability of ICTP when this
work began. O.L.G.P. was supported by Funda\c{c}\~ao de Amparo \`a Pesquisa 
do Estado de S\~ao Paulo (FAPESP) and  by Conselho Nacional
de Ci\^encia e Tecnologia (CNPq). 
 },
        A. Yu. Smirnov\address{The Abdus Salam International Centre
for Theoretical Physics,   I-34100 Trieste, Italy}
                      \address{Institute
for Nuclear Research of Russian Academy  of Sciences, Moscow 117312,
Russia}
}
       
\begin{document}

\begin{abstract}
We study the possible manifestation of the interference between
the effects produced in the atmospheric neutrinos due to
oscillation driven by the solar parameters
parameters $\Delta m^2_{21}$, $\sin^2 2\theta_{21}$ and due to
oscillation driven by $U_{e3}$. 

\vspace{1pc}
\end{abstract}

\maketitle

\section{Introduction}

Recent results on atmospheric neutrinos~\cite{SK-TAUP2001} 
as well as results from the long base-line  experiment K2K \cite{k2k} 
further confirmed the interpretation of the atmospheric neutrino 
anomaly in terms of  
$\nu_\mu\leftrightarrow \nu_\tau$ oscillations with maximal 
or close to  maximal mixing and mass squared difference in the
interval, $\Delta m^2_{atm} = (1.5 - 4) \times 10^{-3} {\rm eV}^2~, 
\sin^2 2\theta_{atm} > 0.88,$ at 90 \% ~ {\rm  C.L.} . 

A sub-dominant oscillation of electron neutrinos is not excluded yet. 
It seems that there is an excess of the $e-$like events in the
low energy part of the sub-GeV sample ($p < 0.4$ GeV, where $p$ is the
momentum of lepton).  In comparison with predictions  based on  
the atmospheric neutrino flux from Ref.\cite{honda} the excess 
is about (12 - 15)\%. For higher energies, the excess is much smaller. 

Can~these~results~be~related~to~the~$\nu_e-$oscillations? 
What could be the implications of the positive answer? 
We have some preliminary results that we will discuss in next sections.
\begin{figure}[htbp]
\vspace*{-1.4cm}
\centerline{\protect\hbox{\psfig{file=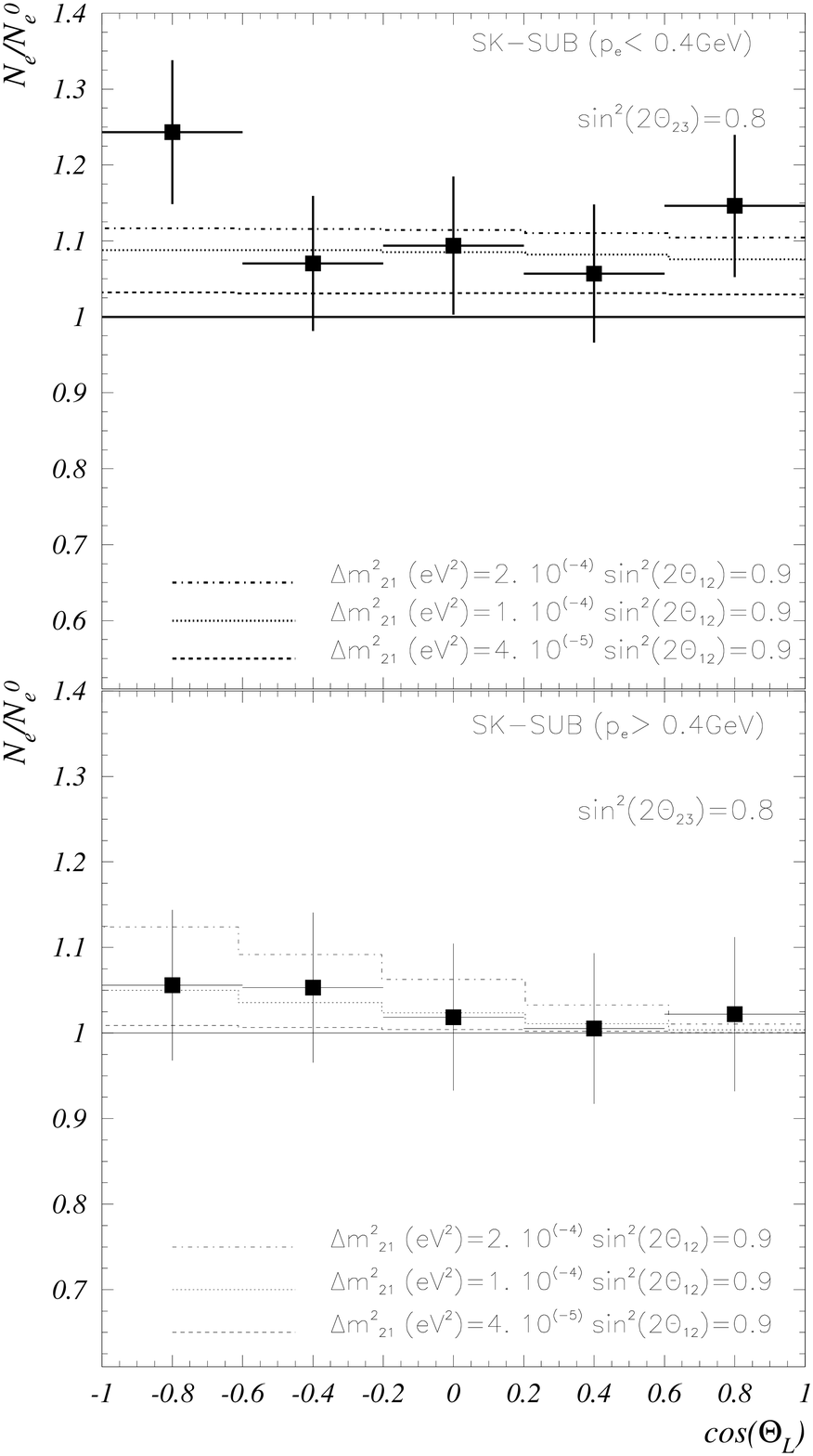,height=12.4cm,width=8.cm}}}
\vskip -1.5cm
\caption{Zenith distribution for sub-GeV events with p$<0.4$ GeV and
for $p>0.4$~GeV. We assume the parameters showed in the plot. }
\label{fig:1}
\end{figure}

\section{Formalism}

In the three neutrino schemes which explain the atmospheric and solar 
neutrino data, there are two possible channels of the 
$\nu_e-$ oscillations:

1.  $\nu_e-$oscillations driven by $\Delta m^2_{atm}$ 
responsible for dominant mode of the atmospheric 
neutrino oscillations~\cite{ADLS}.
These oscillations require non-zero value of $U_{e3}$. 
The effects are restricted by the 
CHOOZ result~\cite{CHOOZ}.

2.  $\nu_e-$oscillations driven by the solar mass splitting 
$\Delta m^2_{\odot}$~\cite{orl1}.

The detailed study of the effect have been done in  our previous  paper
\cite{orl1} where we have  shown that neutrino oscillations with  
parameters in the LMA MSW allowed region~\cite{solar} 
$\Delta m^2_{\odot} =  (2 - 30) \cdot 10^{-5}$ eV$^2$, 
$\sin^2 2\theta_{\odot} > 0.65$, favored by analyzes of solar neutrino
data from SNO~\cite{sno} and
Super-Kamiokande~\cite{sk-sol} 
data, can  lead to an observable excess of the 
e-like events in the sub-GeV atmospheric neutrino sample. 

It was shown that the excess is determined by the 
two neutrino transition probability $P_2$ and  the ``screening" factor:   
\begin{equation}
\frac{F_e}{F_e^0} - 1 =  P_2(r c_{23}^2 - 1)~,
\label{fluxe}
\end{equation}
where $F_e$ and $F_e^0$ are the electron neutrino fluxes with and 
without oscillations and $r$ is the ratio of the original muon and
electron neutrino fluxes. In the sub-GeV region $r \approx 2$,  so 
that the screening factor is zero when the $\nu_{\mu} - \nu_{\tau}$  
mixing is maximal. We show in Fig.~\ref{fig:1} our previous results
compared with the latest data on Super-Kamiokande~\cite{SK-TAUP2001}. 

In previous studies the effects of oscillations 
driven by the solar  and atmospheric  $\Delta m^2$ have been considered 
separately: The studies of the $\Delta m^2_{atm}-$driven 
oscillations  where performed in the framework of the so called ``one level
dominating scheme''  when the effect of solar 
mass splitting between two lightest states,  
$\Delta m^2_{21}$, is neglected. In  studies of the solar $\Delta
m^2_{21}$ driven oscillations it was assumed that  $U_{e3}$ is
negligible.

In this paper  we study the effects of the interplay of 
oscillations with the LMA parameters and non-zero $U_{e3}$. 

\section{$U_{e3}$ and induced interference}

We consider the three-flavor neutrino system  with  
hierarchical mass squared differences: 
$\Delta m^2_{21} = \Delta m^2_{\odot}<< \Delta m^2_{31} = \Delta
m^2_{atm}$.
The evolution of the neutrino vector of state $\nu_f \equiv (\nu_e,
\nu_{\mu}, \nu_{\tau})^T$ is described by the equation 
\begin{equation}
i \frac{d \nu_f}{dt} =
\left( \frac{U M^2 U^\dagger}{2 E} + V \right) \nu_f, 
\label{evolution}
\end{equation} 
where $E$ is the neutrino energy and  
$M^2 = diag(0, \Delta m_{21}^2, \Delta m_{31}^2)$
is the diagonal matrix of neutrino mass squared eigenvalues. 
$V = diag(V_e, 0 ,0)$ is the matrix of matter-induced neutrino potentials
with $V_e = \sqrt 2 G_F N_e$, $G_F$ and $N_e$ being the Fermi constant 
and the electron number density, respectively.   
The mixing matrix $U$ is defined through $\nu_f = U \nu_{mass}$, where
$\nu_{mass} \equiv 
(\nu_1, \nu_2, \nu_3)^T$ is the vector of neutrino mass
eigenstates. It can  be parameterized as 
$
U = U_{23} U_{13} U_{12}.   
$
The matrix $U_{ij}= U_{ij}(\theta_{ij})$ performs the rotation  
in the $ij$- plane by the angle $\theta_{ij}$. 
Here we have neglected possible CP-violation effects in the lepton sector.  
 
\subsection{Propagation basis}

The dynamics of oscillations is simplified in the   
``propagation" basis 
$\tilde{\nu} = (\tilde{\nu}_e, \tilde{\nu}_{2}, \tilde{\nu}_{3})^T$,   
which is related with the flavor basis by  
$\nu_f = \tilde{U} \tilde{\nu}$.
We define the propagation basis in such a way that projection matrix 
$\tilde{U}$ equals:  $\tilde{U} = U_{23} U_{13}$~. 
The propagation basis can be
introduced in
the following way. First, let us  perform the rotation 
$\nu_f =  U_{23} U_{13} \nu'$. Using Eq.~(\ref{evolution}) we find that 
in the basis $\nu'$ the Hamiltonian takes the form,
\begin{equation}
H' \approx
\left(\begin{array}{cc}
H_2   & 0 \\
0  & \Delta m_{31}^2/2E  + V_e s_{13}^2
\end{array}\right)\, ,
\label{matr2}
\end{equation}
where $H_2  = U_{12} M_2  U_{12}^{\dagger} / 2E + V_e c_{13}^2$, 
and $ M_2 = diag(0, \Delta m_{21}^2)$.   
We neglect off-diagonal terms in the evolution equation, 
Eq. (\ref{evolution}).

The evolution matrix S in the propagation basis 
$(\tilde{\nu}_e, \tilde{\nu}_{\mu}, \tilde{\nu}_{\tau})$   
has the following form:
\begin{equation}  
\tilde{S} \approx 
\left(\begin{array}{ccc} 
\tilde{A}_{ee}   & \tilde{A}_{e \mu}    & 0 \\
\tilde{A}_{\mu e}   & \tilde{A}_{\mu \mu}    &   0    \\
0        & 0         & \tilde{A}_{\tau \tau} 
\end{array}
\right) ~~ ,~~
\label{matr-s}
\end{equation}
where $A_{\tau \tau} \approx  \exp(-i\Delta m_{31}^2 L/ 2E)\,, $
and  $L$ is the total distance traveled by the neutrinos. 
\begin{figure}[htbp]
\vspace*{-3.8cm}
\centerline{\protect\hbox{\psfig{file=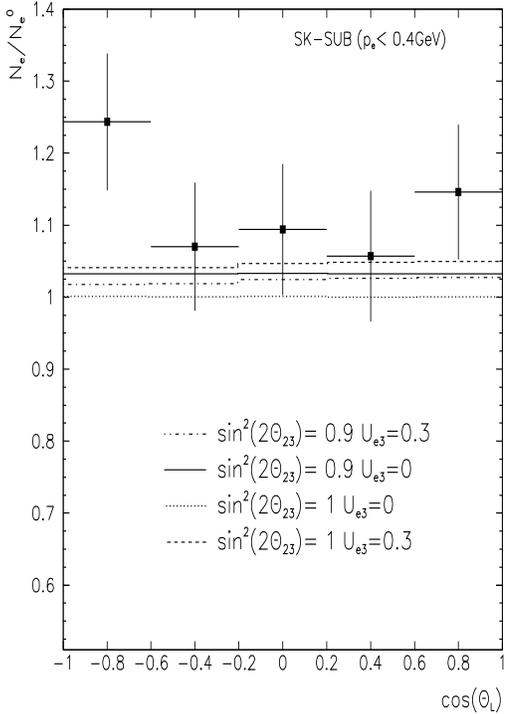,height=12.4cm,width=8.9cm}}}
\vskip -2.6cm
\caption{Zenith distribution for sub-GeV events with p$<0.4$ GeV. 
We assume the parameters showed in the plot and 
$\Delta m_{21}^2=5\,\,10^{-5}$ eV$^2$. }
\label{fig:2}
\end{figure}

\subsection{Flavor transitions and interference}

Let us find  the  probabilities of  
($\nu_{\mu} \leftrightarrow \nu_e$) oscillations, $P_{\mu e}$,  and 
($\nu_e \leftrightarrow \nu_e$) oscillations,  
$P_{ee}$,  relevant for our problem. 
The $S-$matrix  in the flavor basis 
equals: $S = \tilde{U} \tilde{S} \tilde{U}^{\dagger}$,     
and we find 
\begin{eqnarray}
P_{\mu e} = \left| 
-s_{13} c_{13} s_{23} \tilde{A}_{ee} + 
c_{13} c_{23} \tilde{A}_{\mu e} 
\right|^2  
+  s_{13}^2 c_{13}^2 s_{23}^2 \nonumber ,
\label{mue-pr}
\end{eqnarray}
and $P_{ee} = c_{13}^4 (1 - \tilde{P}_{\mu e}) + s_{13}^4$. 
For sub-GeV sample oscillations driven by $\Delta m_{31}^2$ 
are averaged out, so that there is no interference effect due to 
state $\tilde{\nu}_{\tau}$. At the same time, 
according to (\ref{mue-pr}) 
the amplitudes $\tilde{A}_{ee}$ and $\tilde{A}_{\mu e}$ 
interfere. It this interference  which produces effect
we are interested in this paper. Notice that amplitudes
$\tilde{A}_{ee}$ and $\tilde{A}_{\mu e}$   
are both due to solar oscillation parameters. 
However their interference appears 
due to presence  of the third neutrino (non-zero $s_{13}$). 
In what follows we will call the interference of the amplitudes 
(with solar oscillation parameters) due to non-zero $U_{e3} \sim s_{13}$ 
as {\it induced interference}.  
 
Combining $P_{\mu e}$ and $P_{ee}$, the  excess of the $\nu_e-$flux equals:
\begin{eqnarray}
\frac{F_e}{F_e^0} -  1 =  (r c_{23}^2 - 1) \tilde{P}_{\mu e} 
- r s_{13} c_{13}^2 \sin 2\theta_{23} Q \quad\quad\quad\quad\quad
 & & \nonumber \\  
-s_{13}^2 \left[ 2 W_{23} +\tilde{P}_{\mu e} (r - 2) \right] 
+ s_{13}^4 W_{23} (2 - \tilde{P}_{\mu e}) \quad\quad\quad & & \nonumber
\label{fl-excess} 
\end{eqnarray}
and  $Q\equiv Re(\tilde{A}_{ee}^*\tilde{A}_{\mu e})$ 
and $W_{23}\equiv (1 - r s_{23}^2)$. 
The first term on the left hand side (zero order in
$s_{13}^2$) corresponds 
to the contribution we have discussed
in~\cite{orl1}. 
The second term is the effect of the induced interference. 
Let us stress its properties: 
1). The interference term  depends on $s_{13}$ linearly. 
So its effect may not be strongly suppressed even for small 
$s_{13}$. The interference depends  on the  sign of $s_{13}$, also 
does not have  screening factor, 
and its smallness is mainly due to smallness of $s_{13}$.
3). Beside this term has opposite signs for neutrinos and 
anti-neutrinos.

We have calculated  dependences of the excess of
the $e-$like events on the zenith angle of electron, $\Theta_e$. The
procedure was described before in Ref.~\cite{compute}. In Fig.~2  we
show the zenith angle dependences of 
the excess of the e-like  events for  
different values of oscillation parameter.

Concluding, we show that if the LMA solution is the correct one 
and for $\theta_{23}=45^\circ$ ( in this case the effects due the oscillations
driven by $\Delta m_{21}^2$ only are suppressed) we can have a
direct way to determine $U_{e3}$, from the electron neutrino zenith
distribution as is shown  in Figure 2.

\end{document}